\documentstyle[aps,multicol,epsf]{revtex}
\tightenlines
\begin{document}
\draft
\title{Driven Depinning in Anisotropic Media}
\author{Lei-Han Tang\cite{TangAdd},
Mehran Kardar\cite{KardarAdd},
and Deepak Dhar\cite{DharAdd}}
\address{Issac Newton Institute for Mathematical Sciences,
Cambridge University, Cambridge CB3 OEH, England}
\date{August, 1994}
\maketitle
\begin{abstract}
We show that the critical behavior of a driven interface,
depinned from quenched random impurities, depends on the
isotropy of the medium. In anisotropic media the interface
is pinned by a bounding (conducting) surface characteristic
of a model of mixed diodes and resistors.
Different universality classes describe depinning along
a hard and a generic direction. The exponents in the latter
(tilted) case are highly anisotropic, and obtained
exactly by a mapping to growing surfaces.
Various scaling relations are proposed in the former case
which explain a number of recent numerical observations.

\end{abstract}

\begin{multicols}{2}

The pinning of interfaces by impurities occurs in many circumstances
such as in random magnets or fluid flow through porous media.
There has been considerable recent progress in understanding
such collective depinning phenomena.
Insights gained from charge density waves\cite{NF92}
have been extended to describe the critical behavior
of depinning interfaces\cite{NSTL,NF93}.
The renormalization group (RG) analysis indicates that
the interface is a self-affine fractal at the depinning transition.
Narayan and Fisher have argued that the
roughness exponent $\zeta$, of a $d$-dimensional critical
interface is $(4-d)/3$,
to all orders of perturbation theory\cite{NF93}.
However, a number of numerical\cite{Rob,TL92,Boston92}
and experimental results\cite{EXP,Boston92}, mostly in $d=1$,
have cast doubt on the generality of this consclusion.

Amaral, Barabasi, and Stanley\cite{ABS} (ABS) have observed that
numerical results fall roughly into two groups, which they
classify according to the dependence of
the average interface velocity $v(s)$ on its slope $s$.
In one class, the slope dependence is either absent or {\it vanishes}
at the threshold. In the other, $\lambda_{\rm eff}\equiv v''(0)$
{\it diverges} on approaching the depinning transition.
We suggest that a more natural
classification is obtained by considering the dependence of the
threshold force $F_c(s)$ on slope;  in turn related
to the anisotropy of the random medium. The importance of such
slope dependence, and the role of anisotropy,
has been hinted at in a number of recent
publications\cite{NF93,Parisi,LT93,CHV,Semjon93,Zhang94}, but we believe
that it has not  been clearly elucidated.
As a bonus, we find a third (and new) universality class
describing the depinning of interfaces tilted with respect to
the anisotropy axis. Interestingly, by taking advantage of a
mapping to  growing surfaces in one lower dimension, we can {\it exactly}
calculate the highly anisotropic roughness exponents of such
tilted surfaces. The results are confirmed by numerical simulations
in one and two dimensions.

Theoretical studies of interface depinning usually start with the
continuum equation,
\begin{equation}
{\partial {h({\bf x},t)}\over \partial t}=\nabla^2 h
+F+f({\bf x},h),
\label{qEW}
\end{equation}
where $h({\bf x},t)$ is the height of the interface at position ${\bf x}$
at time $t$. The first term on the right hand side describes the smoothening
effect of surface tension, the second term the uniform driving force,
and the third a random force with short range correlations.
This equation arises naturally from the energetics of a domain-wall
in a disordered medium close to equilibrium\cite{Bruinsma};
its applicability to describing
fluid flow in a porous medium\cite{KL} is less well-justified.
Far from equilibrium, the most relevant local term consistent
with translational symmetry is $\lambda(\nabla h)^2/2$.
The usual mechanisms for generating such a term are of kinematic origin
\cite{KPZ} ($\lambda\propto v$) and can be shown to be
irrelevant at the depinning threshold where the velocity $v$
goes to zero\cite{NF93}.  However, if $\lambda$
{\it is not} proportional to $v$ and stays finite at the transition,
it is a relevant operator and expected to modify the critical behavior.
As we shall argue below, anisotropy in the medium is a possible source
of the nonlinearity at the depinning transition.

A model flux line (FL) confined to move in a plane\cite{Dong,TFG} provides an
example where both mechanisms for the nonlinearity are present.
Only the force normal to the FL is responsible for motion,
and is composed of three components: {\bf (1)} A term proportional to
curvature arising from the smoothening effects of line tension.
{\bf (2)} The Lorentz force due to a uniform current density perpendicular to
the plane acts in the normal direction and has a uniform magnitude
$F$ (per unit line length). {\bf (3)} A random force
${\hat {\bf n}}\cdot {\bf f}$
due to impurities, where ${\hat {\bf n}}$ is the unit normal vector\cite{TFG}.
Equating viscous dissipation with the work done by the normal force leads
to the equation of motion
\begin{equation}
{\partial h\over \partial t}=\sqrt{1+s^2}\left[
{\partial_x^2 h\over (1+s^2)^{3/2}}+F+{f_h-sf_x \over \sqrt{1+s^2}}\right],
\label{F-L}
\end{equation}
where $h(x,t)$ denotes transverse displacement of the line and
$s\equiv\partial_x h$. The  nonlinearities generated
by $\sqrt{1+s^2}$ are kinematic in origin\cite{KPZ} and irrelevant
as $v\to0$\cite{NF93}, as can be seen easily by taking
them to the left hand side of Eq.(\ref{F-L}). The shape of the pinned
FL is determined entirely by the competition of the terms in the
square brackets. Although there is no explicit simple $s^2$ term in this
group, it will be generated if the system is {\it anisotropic}.

To illustrate the idea, let us take $f_h$ and $f_x$ to be independent
random fields with amplitudes $\Delta_h^{1/2}$ and $\Delta_x^{1/2}$
respectively; each correlated isotropically in space within
a distance $a$. For weak disorder, a deformation of order $a$
in the normal direction $\hat{\bf n}$ takes place over a distance
$L_c\gg a$ along the line. The total force due to curvature on this piece
of the line is of the order of $L_c(a/L_c^2)$, and the pinning force,
$[(L_c/a)(n_h^2\Delta_h+n_x^2\Delta_x)]^{1/2}$.
Equating the two forces\cite{Bruinsma} yields
$L_c=a(n_h^2\Delta_h+n_x^2\Delta_x)^{-1/3}$ and
an effective pinning strength per unit length,
\begin{equation}
F_0(s)=aL_c^{-2}=a^{-1}\Bigl({\Delta_h+s^2\Delta_x\over 1+s^2}\Bigr)^{2/3}.
\label{F0}
\end{equation}
The roughening by impurities thus reduces
the effective driving force
on the scale $L_c$ to $\tilde F(s)=F-F_0(s)$.
Therefore, even if initially $F$ is independent of $s$, such
a dependence is generated under coarse graining,
{\it provided that the random force
is anisotropic}, i.e. $\Delta_h\ne\Delta_x$.
An expansion of $\tilde F(s)$ around its maximum (which defines the
hard direction) yields an $s^2$ term which is positive and
remains finite as $v\rightarrow 0$.

The FL indicates the origin of the two types of behavior for
$\lambda_{\rm eff}=v''(s=0)$
observed by ABS\cite{ABS}: Nonlinearities of
kinematic origin are proportional to $v$ and vanish at the threshold;
those due to anisotropy survive (and diverge) at the depinning transition.
An immediate consequence of the latter is that the
depinning threshold $F_c$ depends on the average orientation of the line.
In addition, due to the relevance of this term in the RG sense
for $d\leq 4$, the critical behavior at the transition
is modified. A one-loop RG of Eq.(\ref{qEW}) with the
added nonlinearity was carried out by Stepanow\cite{Semjon93}.
He finds no stable fixed point for $2\leq d\leq 4$, but his numerical
integration of the one loop RG equations in $d=1$ yield
$\zeta\approx 0.8615$ and a dynamical exponent $z=1$.
Due to the absence of Galilean invariance,
there is also a renormalization of $\lambda$ which is related
to the diverging $\lambda_{\rm eff}$ observed in
Ref. \cite{ABS}.
The nonperturbative nature of the fixed point precludes a
gauge of the reliability of these exponents.

Numerical simulations of Eq.(\ref{qEW}), {\it with an added
$(\nabla h)^2/2$}
in $d=1$\cite{CHV,Zhang94}, indicate that it shares the characteristics
of a class of lattice models\cite{TL92,Boston92} where the external force
is related to the density $p$ of  `blocking sites' by $F=1-p$.
When $p$ exceeds a critical value of $p_c$, blocking sites form
a directed percolating path which stops the interface.
For a given geometry, there is a direction along which the first
spanning path appears. This defines a {\it hard} direction
for depinning where the threshold force $F_c(s)$ reaches maximum.
Higher densities of blocking sites are needed to form a spanning
path away from this direction, resulting in a lower
threshold force $F_c(s)$ for a tilted interface.
Thus on a phenomenological level we believe that Eq.(\ref{qEW})
modified by the inclusion of nonlinearity,
and directed percolation (DP) models of interface depinning
belong to the same universality class of {\it anisotropic depinning}.
This analogy may in fact be generalized
to higher dimensions, where the blocking path
is replaced by a directed blocking surface\cite{Dhar81,Buldyrev}.
Unfortunately, little is known analytically about the scaling properties
of such a surface at the percolation threshold.

As emphasized above, the hallmark of anisotropic depinning
is the dependence of the threshold force $F_c(s)$ on the
slope $s$.  Above this threshold, we expect
$v(F,s)$ to be an analytical function of $F$ and $s$.
In particular, for $F>F_c(s)$, there is a small $s$ expansion
$v(F,s)=v(F,s=0)+\lambda_{\rm eff}s^2/2+\cdots$.
On the other hand, we can associate a characteristic slope
$\overline{s}=\xi_\perp/
\xi_\parallel\sim(\delta F)^{\nu (1-\zeta)}$, to DP clusters
where $\delta F=F-F_c(0)$, and $\nu $ is the correlation length
exponent. Scaling then suggests
\begin{equation}
v(F,s)=(\delta F)^{\theta}g(s/\delta F^{\nu (1-\zeta)}),
\label{velocity}
\end{equation}
where  $\theta=\nu (z-\zeta)$. Matching Eq.(\ref{velocity}) with
the small $s$ expansion, we see that $\lambda_{\rm eff}$ diverges
as $(\delta F)^{-\phi}$ (as defined by ABS\cite{ABS}) with
$\phi=2\nu (1-\zeta)-\theta=\nu (2-\zeta-z)$.
In $d=1$, the exponents $\nu $ and $\zeta$ are
related to the correlation length exponents $\nu_\parallel$ and $\nu_\perp$
of DP\cite{DP} via $\nu =\nu_\parallel\approx 1.73$ and
$\zeta=\nu_\perp/\nu_\parallel\approx 0.63$,
while the dynamical exponent is $z=1$.
Scaling thus predicts $\phi\approx 0.63$, in agreement with
the numerical result of $0.64\pm 0.08$ in Ref.\cite{ABS}.
Close to the line $F=F_c(0)$ (but at a finite $s$), the dependence
of $v$ on $\delta F$ drops out and we have
\begin{equation}
v(F_c,s)\propto |s|^{\theta/ \nu (1-\zeta)}.
\label{vFc}
\end{equation}
As $z=1$ in $d=1$, the above equation
reduces to $v\propto |s|$, in agreement with Fig.(1) of Ref.\cite{ABS}.
Note that Eqs.(\ref{vFc}) and (\ref{Fc}) are valid also in higher
dimensions, though values of the exponents quoted above
vary with $d$\cite{Buldyrev}. As $F=F_c(s)$ is the line where
$v(F,s)$ vanishes, Eq.(\ref{velocity}) suggests
\begin{equation}
F_c(s)-F_c(0) \propto -|s|^{1/\nu (1-\zeta)}.
\label{Fc}
\end{equation}

An interface tilted away from the hard direction not only
has a different depinning threshold, but also completely different
scaling behavior at its transition. This is because,
due to  the presence of an average interface gradient
${\bf s}=\left\langle \nabla h \right\rangle$, the isotropy
in the internal ${\bf x}$ space is lost. The equation of motion
for fluctuations, $h'({\bf x},t)=h({\bf x},t)-{\bf s}\cdot{\bf x}$, around the
average interface position may thus include terms such as
$\kappa{\bf s}\cdot\nabla h'$, which break the rotational symmetries
in ${\bf x}$ space.  The resulting depinning transition belongs
to yet a new universality class with {\it anisotropic}
response and correlation functions in directions parallel
and perpendicular to ${\bf s}$; i.e.
\begin{eqnarray*}
\left\langle [h({\bf x})-h({\bf x'}) ]^2\right\rangle
&=&|x_\parallel-x_\parallel'|^\zeta
{\cal F}\left({|{\bf x_t}-{\bf x'_t}|\over|x_\parallel-x_\parallel'|^\eta}
\right)\\ \nonumber
&\to&\cases{
|x_\parallel-x_\parallel'|^\zeta& for ${\bf x_t}-{\bf x'_t}=0$\cr
|{\bf x_t}-{\bf x'_t}|^{\zeta/\eta}& for $x_\parallel-x_\parallel'=0 $},
\end{eqnarray*}
where $\eta$ is the {\it ansiotropy} exponent, and ${\bf x}_t$
denotes the $d-1$ directions transverse to ${\bf s}$.

A suggestive mapping allows us to determine the exponents for
depinning a tilted interface: Imagine pushing up all points
on the interface along a $(d-1)$-dimensional cross section of
fixed $x_\parallel$. This move decreases the slope of the
interface uphill but increases it downhill.
Since $F_c(s)$ decreases with increasing $s$,
at criticality the perturbation propagates only a finite
distance uphill but causes a downhill avalanche.
The disturbance front moves at a constant
velocity ($\delta x_\parallel\propto t$) and hence $z_\parallel=1$.
Furthermore, the evolution of successive cross sections
${\bf x}_t(x_\parallel)$ is expected to be the same as the
evolution in time of a $(d-1)$-dimensional interface! The latter is
governed by the Kardar-Parisi-Zhang (KPZ) equation\cite{KPZ},
whose scaling behavior has
been extensively studied. From this analogy we conclude,
\begin{equation}
\zeta(d)={\zeta_{\rm KPZ}(d-1)\over z_{\rm KPZ}(d-1)},\quad
\eta(d)={1\over z_{\rm KPZ}(d-1)}.
\label{exponent_kpz}
\end{equation}
In particular, the tilted interface with
$d=2$ maps to the growth problem in 1+1 dimensions where the exponents
are known exactly, yielding $\zeta(2)=1/3$ and $\eta(2)=2/3$.
This picture can be made more precise for a lattice model introduced below.
Details will be presented elsewhere.

To get the exponent $\theta$ for the vanishing of velocity
of the tilted interface, we note that since $z_\parallel=1$,
$v$ scales as the excess slope $\delta s=s-s_c(F)$.
The latter controls the density of the above moving fronts;
$s_c(F)$ is the slope of the critical interface at a given driving force $F$,
i.e., $F=F_c(s_c)$.
Away from the symmetry direction,
the function $F_c(s)$ has a non-vanishing derivative and hence
\begin{equation}
\delta F=F-F_c(s)=F_c(s_c)-F_c(s)\sim \delta s\sim v.
\label{v_tilt}
\end{equation}
We thus conclude that generically $\theta=1$ for tilted interfaces,
independent of dimension.

Due to scarcity of analytical results, there is need for a simple model
suitable for numerical investigation. We propose a
variant of previously studied percolation models of interface
depinning\cite{TL92,Buldyrev}
with the essential ingredient of a slope dependent threshold.
A solid-on-solid (SOS) interface is described by a set of integer heights
$\{h_{\bf i}\}$
where ${\bf i}$ is a group of $d$ integers. With each configuration is
associated a random set of
pinning forces  $\{\eta_{\bf i}\in[0,1)\}$. The heights are updated
{\it in parallel} according to the following rules:
$h_{\bf i}$ is increased by one if (i)
$h_{\bf i}\leq h_{\bf j}-2$ for at least one
${\bf j}$ which is a nearest neighbor of ${\bf i}$, {\it or} (ii)
$\eta_{\bf i}<F$ for a pre-selected uniform force $F$. If
$h_{\bf i}$ is increased, the associated random force $\eta_{\bf i}$ is also
updated, i.e. replaced by a new random number in the interval $[0,1)$.
Otherwise, $h_{\bf i}$ and $\eta_{\bf i}$ are unchanged.
The simulation is started with initial conditions
$h_{\bf i}(t=0)={\rm Int}[s {\bf i}_x]$, and
boundary conditions $h_{\bf i+L}={\rm Int}[s L]+h_{\bf i}$ are
enforced throughout.
The CPU time is greatly
reduced by only keeping track of active sites.

The above model has a simple analogy to
a resistor-diode percolation problem\cite{Dhar81,Buldyrev}.
Condition (i) ensures that, once a site $({\bf i},h)$ is wet (i.e.,
on or behind the interface),
all neighboring columns of {\bf i} must be wet
up to height $h-1$. Thus there is always `conduction' from a site
at height $h$ to sites in the neighboring columns at height $h-1$.
This relation can be represented by diodes pointing diagonally downward.
Condition (ii) implies that `conduction' may also occur upward.
Hence a fraction $F$ of vertical bonds are turned into resistors
which allow for two-way conduction. Note that, due to the SOS
condition, vertical downward conduction is always possible.
For $F<F_c$, conducting sites connected to a point lead
at the origin, form a cone whose hull is the interface separating
wet and dry regions. The opening angle of the cone
increases with $F$, reaching $180^\circ$ at $F=F_c$,
beyond which percolation in the entire space
takes place, so that all sites are eventually wet.
If instead of a point, we start with a planar lead defining
the initial surface, the percolation threshold depends
on the surface orientation, with the highest threshold
for the untilted one.

Our simulations of lattices of $65 536$ sites in $d=1$ and of $512\times 512$
and $840\times 840$ sites in $d=2$ confirm the exponents for depinning
in the hard direction as summarized in Ref.\cite{Buldyrev}.
For a tilted surface in $d=1$ the roughness exponent determined
from the height-height correlation function is consistent with
the predicted value of $\zeta={1/ 2}$ and different from
$\zeta\approx 0.63$ of the untilted one.
The dependence of the depinning threshold on slope is clearly seen
from Fig. 1, where the average velocity is plotted against the driving
force for $s=0$ (open) and $s=1/2$ (solid).
The $s=0$ data can be fitted to a power-law
$v\sim (F-F_c)^\theta$, where $F_c\approx 0.461$, $\theta=0.63\pm 0.04$
for $d=1$, and $F_c\approx 0.201$, $\theta=0.72\pm 0.04$ for $d=2$.
Data at $s={1/ 2}$ is consistent with Eq.(\ref{v_tilt})
close to the threshold.

We also measured height-height correlation functions
at the depinning transition. For a tilted surface in $d=2$,
the height fluctuations and corresponding dynamic behaviors are different
parallel and transverse to the tilt.
Figure 2 shows a scaling plot of
(a) $C_\parallel(r_\parallel,t)\equiv
\langle [h(x_\parallel+r_\parallel,x_t,t)-h(x_\parallel,x_t,t)]^2\rangle$
and (b)
$C_t(r_t,t)\equiv\langle [h(x_\parallel,x_t+r_t,t)-h(x_\parallel,x_t,t)]^2
\rangle$
against the scaled distances at the depinning threshold of
an $s={1/2}$ interface.
Each curve shows data at a given $t=32$, 64, $\cdots$, 1024, averaged
over 50 realizations of the disorder.
The data collapse is in agreement with the mapping to the KPZ equation
in one less dimension.

In summary, critical behavior at the depinning of an interface depends
on the symmetries of the underlying medium. Different universality
classes can be distinguished from the dependence of the threshold
force (or velocity) on the slope, which is reminiscent of similar
dependence in a model of  resistor-diode percolation. In addition to isotropic
depinning,  we have so far identified two classes of anisotropic depinning:
along a (hard) axis of inversion symmetry in the plane,
and tilted away from it.
We have no analytical results in the former case, but suggest a
number of scaling relations that are validated by simulations.
In the latter (more generic) case we have obtained {\it exact}
information from a mapping to moving interfaces, and confirmed
them by simulations in $d=1$ and $d=2$. As it is quite common
to encounter anisotropy for flux lines in superconductors, domain
walls in magnets, and interfaces in porous media, we expect
our results to have important experimental ramifications.

We have benefitted from discussions with D.~Erta\c s and J. Kert\'esz.
MK is supported by NSF grants DMR-93-03667 and PYI/DMR-89-58061).
LHT is supported in part by the DFG through SFB-341.

\begin{figure}
\narrowtext
\caption{
Average interface velocity $v$ versus the driving force $F$,
for $d=1$, $s=0$ (open circles), $d=1$, $s={1/ 2}$ (solid circles),
$d=2$, $s=0$ (open squares), and $d=2$, $s={1/ 2}$ (solid squares).}
\label{fig1}
\end{figure}
\begin{figure}
\narrowtext
\caption{
Height-height correlation functions (a) along and (b) transverse
to the tilt for an $840^2$ system at different times
$32\leq t\leq 1024$. The interface at $t=0$ is flat;
$d=2$, $s={1/2}$, and $F=0.144$.}
\label{fig2}
\end{figure}

\end{multicols}

\end{document}